\documentclass[psfig]{aa}
\usepackage{graphicx}

\begin{document}

\title {Line ratios for Helium-like ions: Applications to collision-dominated plasmas}
\author{ 
        D. Porquet\inst{1,2}
        \and
        R. Mewe\inst{3}
        \and
        J. Dubau\inst{4,5}
        \and
        A.J.J. Raassen\inst{3,6}              
        \and
        J.S. Kaastra\inst{3}
        }
\offprints {Delphine Porquet, \email dporquet@cea.fr}
\institute{
 CEA/DSM/DAPNIA, Service d'Astrophysique, CEA Saclay, 91191 Gif sur Yvette Cedex, France 
 \and
  DAEC, Observatoire de Paris, Section Meudon, 92195 Meudon Cedex, France
 \and
 Space Research Organization Netherlands (SRON),
 Sorbonnelaan 2, 3584 CA Utrecht, The Netherlands
  \and
LSAI, U.M.R. 8624, CNRS, Universit{\'e} de Paris Sud, 91405 Orsay Cedex, France
  \and
  DARC, Observatoire de Paris, Section Meudon, 92195 Meudon Cedex, France
  \and
 Astronomical Institute "Anton Pannekoek", Kruislaan 403,
 1098 SJ Amsterdam, The Netherlands 
} 

\date{Received date 03 May 2001; accepted date 21 June 2001}
\titlerunning{Line ratios for He-like ions (collision-dominated plasmas)}
\authorrunning{Porquet et al.}

\onecolumn
\abstract{
The line ratios $R$ and $G$ of the three main lines of He-like ions (triplet: {\it resonance}, 
{\it intercombination}, 
{\it forbidden} lines) are calculated for \ion{C}{v}, \ion{N}{vi}, \ion{O}{vii}, 
\ion{Ne}{ix}, \ion{Mg}{xi}, 
and \ion{Si}{xiii}. These ratios can be  used to derive electron density 
$n_{\mathrm e}$ and temperature $T_{\mathrm e}$ 
of hot late-type stellar coronae and O, B stars from high-resolution spectra obtained with 
{\sl Chandra (LETGS, HETGS)} and {\sl XMM-Newton (RGS)}. All excitation and radiative processes between the 
levels  and the effect of upper-level cascades from collisional electronic excitation 
and from dielectronic and radiative 
recombination have been considered. When possible the best experimental values for radiative transition 
probabilities are used.  For the higher-Z ions (i.e. \ion{Ne}{ix}, \ion{Mg}{xi}, \ion{Si}{xiii})  possible 
contributions from blended dielectronic satellite lines to each line of the triplets were included in the 
calculations of the line ratios $R$ and $G$ for four specific spectral resolutions: 
{\sl RGS}, {\sl LETGS}, {\sl HETGS-MEG}, {\sl HETGS-HEG}. The influence of an external stellar radiation field on the 
coupling of the 2$^3S$ (upper level of the {\it forbidden} line) and 2$^3P$ levels (upper levels of the 
{\it intercombination} lines) is taken into account. 
This process is mainly important for the lower-Z ions 
(i.e. \ion{C}{v}, \ion{N}{vi}, \ion{O}{vii}) at moderate radiation temperature (T$_{\rm rad}$). 
 These improved calculations were done for plasmas in collisional ionization equilibrium, 
but will be later extended to 
photo-ionized plasmas and to transient ionization plasmas.
The values for $R$ and $G$ are given in extensive tables\footnote{{\small Tables 4 to 69 and A.1 to A.6 are only available in electronic form
at the CDS
 via anonymous ftp to cdsarc.u-strasbg.fr (130.79.128.5)
 or via http://cdsweb.u-strasbg.fr/cgi-bin/qcat?J/A+A/.}}, 
for a large range of parameters, which could be used 
directly to compare to the observations.
\keywords{X-rays: stars: -- atomic processes -- stars: coronae --  stars: late-type -- stars: activity -- early-type stars: activity} 
}
\maketitle
\twocolumn
\section{Introduction}
The new generation of X-ray satellites (\,{\sl Chandra}, {\sl XMM-Newton}\,) 
enables us to obtain unprecedented high 
spectral resolution and high S/N spectra. The wavelength ranges of the {\sl RGS}\footnote{{\small {\sl RGS}: 
Reflection Grating Spectrometer on board {\sl XMM-Newton} (den Herder et al. \cite{denHerder2001}).}} (6--35 \AA), 
of the {\sl LETGS}\footnote{{\sl LETGS}: 
Low Energy Transmission Grating Spectrometer on board {\sl Chandra} (Brinkman et al. \cite{Brinkman2000}).} (2--175 \AA), 
and of the {\sl HETGS}\footnote{{\sl HETGS}: 
High Energy Transmission Grating Spectrometer on board {\sl Chandra} (Canizares et al. \cite{Canizares2000}).} 
(MEG range: 2.5--31 \AA; HEG range: 1.2--15 \AA )  
contain the Helium-like line ``triplets'' from \ion{C}{v} 
(or \ion{N}{vi} for the {\sl RGS}, and for the {\sl HETGS-HEG}) to \ion{Si}{xiii}. The triplet consists of three close lines: the {\it resonance} 
line, the {\it intercombination} line and the {\it forbidden} line.  The Helium-like triplets provide electron 
density ($n_{\mathrm e} \sim$~10$^8$--10$^{13}$~cm$^{-3}$) as well as electron temperature ($T \sim 1$--10~MK) as first 
shown by Gabriel \& Jordan (\cite{Gabriel69}). The line ratios of these He-like triplets enable also the 
determination of the ionization processes (photo-ionization and/or collisional ionization) which prevail 
in the plasma (Porquet \& Dubau \cite{PorquetDubau2000}, Liedahl \cite{Liedahl99}).\\
\indent The ratios of these lines are already widely used for collisional solar plasma diagnostics (e.g., 
Gabriel \& Jordan \cite{Gabriel69}, Doyle \cite{Doyle80}, Keenan et al. \cite{Keenan87}, 
McKenzie \& Landecker \cite{McKenzie82}). \\
Recently, also theoretical calculations for photo-ionized plasmas or ``hybrid'' plasmas (photo-ionization 
plus collisional ionization) have been  given by Porquet \& Dubau (\cite{PorquetDubau2000})
(hereafter  referred to as {\bf Paper I}). 
Their calculations have been already applied to spectra of Seyfert 
galaxies (e.g. \object{NGC\,5548}, Kaastra et al. \cite{Kaastra2000}; 
\object{Mkn\,3}, Sako et al. \cite{Sako2000}; 
\object{NGC\,4151}, Ogle et al. \cite{Ogle2000};
\object{NGC\,4051}, Collinge et al. \cite{Collinge2001}, etc...).\\
 We  present here calculations of these ratios, from \ion{C}{v} to \ion{Si}{xiii}, which could be applied 
 directly for the first time to {\sl Chandra} and {\sl XMM-Newton} observations of  extra-solar collisional 
 plasmas such as stellar coronae. These calculations have been done to apply an improved model to the density 
 analysis of the {\sl RGS}, the {\sl LETGS} and the {\sl HETGS} spectra of various 
late-type stars such as \object{Capella}, \object{Procyon}, 
 and \object{$\alpha$~Centauri} (e.g., Audard et al. \cite{Audard2001}, Ness et al. \cite{Ness2001a}, 
 Mewe et al. \cite{Mewe2001}) and also to O stars such as \object{$\zeta$ Puppis} (Kahn et al. \cite{Kahn2001}). 
Our model is to 
 be considered as an improvement of various previous calculations for solar plasmas such as done by e.g., 
 Gabriel \& Jordan (\cite{Gabriel69}), Blumenthal et al. (\cite{Blumenthal72}), Mewe (\cite{Mewe72}), 
Mewe \& Schrijver (\cite{Mewe78a}, \cite{Mewe78b}, \cite{Mewe78c}),
 Pradhan \& Shull (\cite{Pradhan81}), Mewe \& Gronenschild (\cite{Mewe81}), 
Mewe et al. (\cite{Mewe85}), and Pradhan (\cite{Pradhan82},  \cite{Pradhan85}). 
The calculations are partly based on recent work by Porquet \& Dubau  (\cite{PorquetDubau2000}). \\
\indent In the next three sections, we introduce the plasma diagnostics and the atomic processes and atomic data taken into 
account in the calculations. In sect.~\ref{sec:results}, we display the results for $R$ and $G$ calculated for four specific spectral resolutions (Full Width at Half Maximum: FWHM): {\sl RGS}, {\sl LETGS}, {\sl HETGS-MEG}, 
{\sl HETGS-HEG}, over a broad range of physical parameters: $n_{\rm e}$, $T_{\rm e}$, 
radiation temperature ($T_{\rm rad}$), 
and radiation dilution factor ($W$).

\section{Plasma diagnostics} \label{sec:diagnostics}

\begin{figure}[h!]
\centering
\resizebox{8cm}{!}{\includegraphics{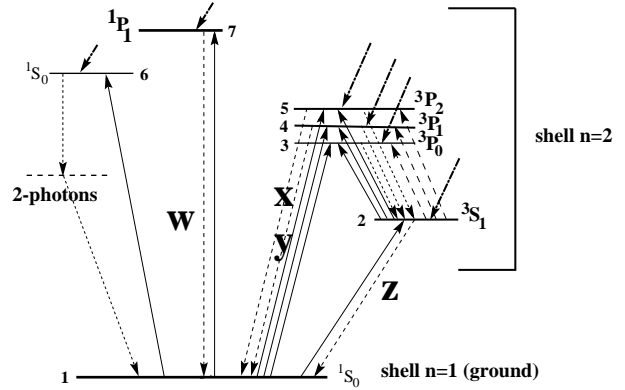}} 
\caption{Simplified level scheme for Helium-like ions. {\bf w} (or r), {\bf x,y} (or i), and {\bf z} (or f): resonance, intercombination, and forbidden lines, respectively. {\it Full upward arrows}: collisional excitation transitions, 
{\it broken arrows}: radiative transitions (including photo-excitation from 2\,$^{3}$S$_{1}$ to 2\,$^{3}$P$_{0,1,2}$ levels, and 2-photon continuum from  2\,$^{1}$S$_{0}$ to the ground level), and  {\it thick skew arrows}:  recombination (radiative and dielectronic) plus cascade processes.}
\label{fig:gotrian}
\end{figure}

\indent In the X-ray range, the three most intense lines of Helium-like ions (``triplet'') are: the 
{\it resonance} line ($w$, also called $r$: 1s$^{2}$\,$^{1}S_{\mathrm{0}}$ -- 1s2p\,$^{1}P_{\mathrm{1}}$), 
the {\it intercombination} lines ($x+y$, also called $i$: 1s$^{2}$\,$^{1}S_{0}$ -- 1s2p\,$^{3}P_{2,1}$) and 
the {\it forbidden} line ($z$, also called $f$: 1s$^{2}$\,$^{1}{S}_{0}$ -- 1s2s\,$^{3}{S}_{1}$). 
They correspond to transitions between the $n$=2 shell and the $n$=1 ground-state shell 
(see Figure~\ref{fig:gotrian}). 
The wavelengths in \AA\, of each line from \ion{C}{v} (Z=6) to \ion{Si}{xiii} (Z=14) 
are reported in Table~\ref{table:lambda}.\\

\begin{table}[!h]
\caption{Wavelengths in \AA\, of the three main X-ray lines of \ion{C}{v}, \ion{N}{vi}, \ion{O}{vii}, 
\ion{Ne}{ix}, \ion{Mg}{xi} and \ion{Si}{xiii} (from Vainshtein \& Safronova \cite{Vainshtein78}).}

\begin{center}
\begin{tabular}{c@{\ }c@{\ }c@{\ }c@{\ }c@{\ }c@{\ }c@{\ }c@{\ }}
\hline
\hline
{\small line}                 &    {\small label}   & {\small \ion{C}{v}}&{\small \ion{N}{vi}}&{\small \ion{O}{vii}}  &{\small \ion{Ne}{ix}} &{\small \ion{Mg}{xi}}  &{\small \ion{Si}{xiii}}\\
\hline
{\small {\it resonance} }         & {\small $w$ ($r$)}&{\small 40.279} &{\small 28.792} &{\small 21.603}  &{\small 13.447} &{\small 9.1681} &{\small 6.6471} \\
{\small {\it inter-}}& {\small $x$ }        &{\small 40.711} &{\small 29.074} &{\small 21.796}  &{\small 13.548} &{\small 9.2267} &{\small 6.6838} \\
{\small {\it combination}}                                       & {\small $y$}        &{\small 40.714} &{\small 29.076} &{\small 21.799}  &{\small 13.551} &{\small 9.2298} &{\small 6.6869} \\
{\small {\it forbidden}}          &{\small  $z$ ($f$)}   &{\small 41.464}  &{\small 29.531}  &{\small 22.095}  &{\small 13.697} &{\small 9.3134} &{\small 6.7394} \\
\hline
\hline
\end{tabular}
\end{center}
\label{table:lambda}
\end{table}

As shown by Gabriel \& Jordan (\cite{Gabriel69}), the ratios defined below are sensitive to the electron density and 
to the electron temperature:
\begin{equation}
R~(n_{\mathrm e})~=~\frac{z}{x+y}~~~~~ \left({\mathrm{also~~}} \frac{f}{i} \right) \\
\label{eq:R}
\end{equation}
\begin{equation}
G~(T_{\mathrm e})=\frac{z+(x+y)}{w}  ~~~~~  \left({\mathrm{also}}~~ \frac{f+i}{r}\right)  \\
\label{eq:G}
\end{equation}

\subsection{Density diagnostic}\label{sec:density}

\begin{figure}[h!]
\centering
\resizebox{8cm}{!}{\includegraphics{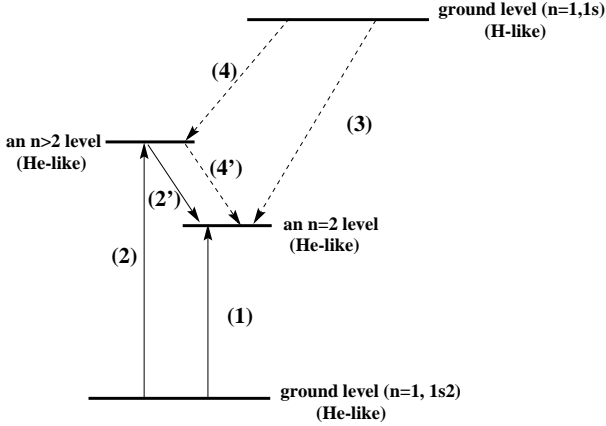}}
\caption{Simplified diagram showing the different contributions to the population of a given $n$=2 shell 
level. (1): direct contribution due to collisional excitation from the ground level (1s$^{2}$) of He-like ions; 
(2)+(2'): collisional excitation plus upper-level radiative cascade contribution; (3): direct radiative 
recombination or direct dielectronic recombination from H-like ions; and (4)+(4'): radiative recombination or 
dielectronic recombination plus upper-level radiative cascade contribution.}
\label{fig:exc_recomb}
\end{figure}

\indent In the low-density limit, all $n$=2 states are populated directly 
or via upper-level radiative cascades by electron impact   
 from the He-like ground state and/or by (radiative and dielectronic) 
recombination of H-like ions (see Figure~\ref{fig:exc_recomb}).
These states decay radiatively directly or by cascades to the ground level. The relative intensities of the 
three intense lines are then independent of density. As $n_{\mathrm{e}}$ increases from the low-density 
limit, some of these states (1s2s\,$^{3}$S$_{1}$ and $^{1}$S$_{0}$) are depleted by collisions to the nearby 
states where $n_{\mathrm{crit}}$\,C\,$\sim$A, with C being the collisional coefficient rate, A being the 
radiative transition probability from $n$=2 to $n$=1 (ground state), and $n_{\mathrm{crit}}$ being the 
critical density. Collisional excitation depopulates first the 1s2s\,\element[][3]{S}$_{1}$ level 
(upper level of the {\it forbidden} line) to the 1s2p\,\element[][3]{P}$_{0,1,2}$ levels (upper levels of 
the {\it intercombination} lines). The intensity of the {\it forbidden} line decreases while those of the 
{\it intercombination} lines increase, hence implying a reduction of the ratio $R$ (according to 
Eq.~\ref{eq:R}), over approximately two or three decades of density (see Fig.~8 in Paper\,I).
For much higher densities, 1s2s\,$^{1}$S$_{0}$ is also depopulated to 1s2p$^{1}$P$_{1}$, and the 
{\it resonance} line becomes sensitive to the density (this has been nicely illustrated by Gabriel \& Jordan 
(\cite{GabrielJordan72}) in their Fig.~4-6-9).\\
 However caution should be taken for low-Z ions (i.e. \ion{C}{v}, \ion{N}{vi}, \ion{O}{vii}) since in case of 
 an intense UV radiation field, the photo-excitation between the $^{3}$S term and the $^{3}$P term is not 
 negligible. This process has the same effect on the {\it forbidden} line and on the {\it intercombination} 
 line as the collisional coupling, i.e. lowering of the ratio $R$, and thus could mimic a high-density plasma. 
 It should be taken into account to avoid any misunderstanding between a high-density plasma and a high 
 radiation field (see $\S$\ref{sec:photo-excitation} for more details).

\subsection{Temperature and ionization process diagnostics}\label{sec:temperature}

\indent The ratio $G$ (see Eq.~\ref{eq:G}) is sensitive to the electron temperature since the collisional 
excitation rates have not the same dependence with  temperature for the {\it resonance} line as for the 
 {\it forbidden} and {\it intercombination} lines.\\
 In addition, as detailed in Paper\,I (see also Mewe \cite{Mewe99}, and Liedahl \cite{Liedahl99}), the relative intensity of the 
 {\it resonance} $w$ (or $r$) line, compared to the {\it forbidden} $z$ (or $f$) and the {\it intercombination } $x+y$ (or $i$) lines, contains 
 information about the ionization processes that occur: a strong  {\it resonance} line compared to the 
 {\it forbidden} or the {\it intercombination} lines corresponds to collision-dominated plasmas. 
It leads to a ratio of $G=(z+(x+y))/w\sim$1 (or $(f+i)/r\sim$1). 
On the contrary, a weak {\it resonance} line corresponds to plasmas dominated 
 by the photo-ionization ($G=(z+(x+y))/w>$4, or $(f+i)/r>$4).

\section{Schematic model}\label{sec:schematic-model}

We illustrate the relevant processes in the formation of the {\it resonance}, {\it intercombination}, and 
{\it forbidden} lines with a simplified level scheme (cf. Mewe \& Schrijver \cite{Mewe78a}) consisting of the 
following levels denoted by short labels: g: ground level 1$^1{\rm S}_0$; m$^{\prime}$: upper metastable 
level 2$^1S_0$ of the two-photon transition; 1$^{\prime}$: upper level 2$^1P_1$ of the {\it resonance} line; m: 
upper metastable level 2$^3S_1$ of the {\it forbidden} ($f$) line; p$_k$ ($k=1,2,3$: levels 
2$^3{\rm P}_k$ (2$^3{\rm P}_1$ is the upper level of the by far the strongest component ($y$) of the 
{\it intercombination} line $i$, and 2$^3{\rm P}_2$ is the upper level of the weaker component ($x$)); c: continuum 
level which lumps together all higher levels to represent the cascades from excitation and recombination 
processes.

The electron collisional rate coefficient (in cm$^3$~s$^{-1}$) for transition $j \to k$ is written as: 
\begin{equation}
C_{jk} = 8.63\ 10^{-6} {\gamma_{jk}\over {w_j \sqrt{T_{\mathrm e}}}} {\rm exp}\Bigl(-{{\Delta E_{jk}}\over {kT_{\mathrm e}}}\Bigr),
\end{equation}
where $\Delta E_{jk}$ is the excitation energy, $T_{\mathrm e}$ is the electron temperature in K, $\gamma_{jk}$ the 
collision strength averaged over a Maxwellian electron energy distribution, and $w_{j}$ the statistical weight of 
the initial level $j$. 

The total rate coefficients for the formation of the {\it forbidden} ($f$ or $z$ notation) 
and {\it intercombination} ($i$ or $x+y$) 
line can be written as (e.g., Mewe \& Schrijver \cite{Mewe78a}, Eqs. 18--30):
\begin{equation}
I_f = BR_{mg} \Bigl[ C_{gm} + \sum_{k=0}^2 BR_{p_km} C_{gp_k}\Bigr],
\end{equation}
\begin{equation}
I_i = \sum_{k=1}^2 BR_{p_kg} \Bigl[ C_{gp_k} + BR_{mp_k} C_{gm} \Bigr],
\end{equation}
with the various branching ratios:
\begin{equation}
BR_{mg} = {A_{mg}\over {A_{mg} + n_{\mathrm e} S_{mp}^{\prime}}},
\end{equation}
\begin{equation}
S_{mp}^{\prime} = \sum_{k=1}^2 C_{mp_k}BR_{p_kg}^{\prime},
\end{equation}
\begin{equation}
BR_{p_km} = {A_{p_km}\over {A_{p_km} + A_{p_kg}}} \ \ (BR_{p_0m} \equiv 1),
\end{equation}
\begin{equation}
BR_{p_kg}^{\prime} = {A_{p_kg}\over {A_{p_kg} + A_{p_km}}} \ \ (BR_{p_0g}^{\prime} \equiv 0),
\end{equation}
\begin{equation}
BR_{p_kg} = {A_{p_kg}\over {A_{p_kg} + A_{p_km}BR_{mg}}} \ \ (BR_{p_0g} \equiv 0),
\end{equation}
\begin{equation}
BR_{mg} = {A_{mg}\over {A_{mg} + n_{\mathrm e} S_{mp}}},
\end{equation}
\begin{equation}
BR_{mp_k} = {{n_{\mathrm e} C_{mp_k}}\over {A_{mg} + n_{\mathrm e} S_{mp}}},
\end{equation}
\begin{equation}
S_{mp} = \sum_{k=0}^2 C_{mp_k}.
\end{equation}
Analogously we can derive 
the total rate coefficients for the formation of the {\it resonance} ($r$ or $w$) line or two-photon radiation 
(2ph) by substituting $m \rightarrow m^{\prime}$, $p,p_k \rightarrow 1^{\prime}$ and performing no summation:
\begin{equation}
I_r = BR_{1^{\prime}g} \Bigl[ C_{g1^{\prime}} + BR_{m^{\prime}1^{\prime}} C_{gm^{\prime}} \Bigr],
\end{equation}
\begin{equation}
I_{2ph} = BR_{m^{\prime}g} \Bigl[ C_{gm^{\prime}} + BR_{1^{\prime}m} C_{g1^{\prime}}\Bigr],
\end{equation}
and changing all $BR$'s etc. appropriately.

We note that the collision coefficients $C_{jk}$ include also the $n > 2$ cascades.
The radiative transition probabilities are tabulated in the Table 2 of Paper\,I and the 
effective collision strengths $\gamma$ in their Tables 9--13 and in their Fig.~4.
We assume in this section that the electron density is so low that the collision de-excitation rate $n_{\mathrm e}C_{p_km}$ can be 
neglected with respect to the spontaneous radiative rate $A_{p_km}$ (e.g., for C V $n_{\mathrm e}C_{p_km} < A_{p_km}$ 
for $n_{\mathrm e} < 10^{15}$~cm$^{-3}$).
This can be easily taken into account by the substitution
\begin{equation}
A_{p_km} \rightarrow A_{p_km} + n_{\mathrm e} C_{p_km}. 
\end{equation}
Further, all collision processes that couple the singlet and triplet system are neglected in this schematic model.
It turns out that at high density the coupling $m,p_k \leftrightarrow m^{\prime},1^{\prime}$ cannot be 
neglected.\\

 In the full calculations used in this work (see $\S$\ref{sec:results}), 
the coupling between the singlet and triplet system has been taken into account, 
as well as the collisional de-excitation.\\

If we take also into account the contribution from radiative and dielectronic recombinations of the 
Hydrogen-like ion we substitute  
\begin{equation}
C_{gk} \rightarrow C_{gk} + \alpha^{\prime}_{ck},
\end{equation}
where $k = m$ or $k = p_k$, respectively, and
\begin{equation}
\alpha^{\prime}_{ck} = (N_H/N_{He}) \alpha_{ck},
\end{equation}
where $\alpha_{ck}$ is the total radiative and dielectronic recombination rate coefficient including cascades 
and $N_H/N_{He}$ is the abundance ratio of Hydrogen-like to Helium-like ions (e.g. taken from Arnaud \& Rothenflug 
\cite{Arnaud85}, Mazzotta et al. \cite{Mazzotta98}). The recombination coefficients are given in Tables 3--8 in Paper I. 
However, for a collision-dominated plasma, the recombination generally gives 
only a minor effect (i.e. far less than few percents), 
but are nevertheless introduced in the full calculations in section~\ref{sec:results}.

Mewe \& Schrijver (\cite{Mewe78a}) took also into account the effect 
of a stellar radiation field (also called photo-excitation) with effective radiation 
temperature $T_{\rm rad}$. This can be done by substituting in the above equations:
\begin{equation}
n_{\mathrm e} C_{mp_k} \rightarrow n_{\mathrm e} C_{mp_k} + B_{mp_k},
\end{equation}
\begin{equation}
n_{\mathrm e} S_{mp} \rightarrow n_{\mathrm e} S_{mp} + B_{mp},
\end{equation}
where $B_{mp} = \sum_{k=0}^2 B_{mp_k}$ and
\begin{equation}
B_{mp_k} = {{W A_{p_km} (w_{p_k}/w_m)}\over  {{\rm exp}\Bigl({{\Delta E_{mp_k}}\over {kT_{\mathrm rad}}}\Bigr) - 1}},
\end{equation}
is the rate (in s$^{-1}$) of absorption $m \rightarrow p_k$ and $W$ is the dilution factor of the radiation 
field (as a special case we take $W={1\over 2}$ close to the stellar surface). 
 We have checked that the radiation field is so low 
that the induced emission rate $B_{p_km}$=$(w_m/w_{p_k})\,B_{mp_k}$ is negligible respect to the 
spontaneous radiative rate $A_{p_km}$. 
Nevertheless, in the full calculations used in this work (see $\S$\ref{sec:results}), 
we have taken into account induced emission as well as photo-excitation.

Mewe \& Schrijver (\cite{Mewe78a}) considered also  inner-shell ionization of the Lithium-like ion which can give
an important contribution to the {\it forbidden} line in a transient plasma (Mewe \& Schrijver \cite{Mewe78b}) such as a supernova remnant.
However, in the present calculations we neglect this because we consider plasmas in ionization equilibrium.

Finally, Mewe \& Schrijver (\cite{Mewe78a},\cite{Mewe78c}) have considered 
also excitation 2$^3$S$ \to $ 2$^3$P by proton collisions
using approximations of Coulomb-Born results from Blaha (\cite{Blaha71}). 
In a new version (SPEX90) of our spectral code SPEX (Kaastra et al. \cite{Kaastra96}), which contains for the 
H- and He-like ions an improvement of the known MEKAL code (Mewe et al. \cite{Mewe85}, \cite{Mewe95a}), proton collisions are
taken into account based on Blaha's results. Test calculations with SPEX90 show that for an equilibrium plasma 
in all practical cases proton excitation is negligible compared to electron excitation. 

In the case where $N_H/N_{He} \gg 1$, recombination is dominant e.g., for photo-ionized plasmas  it turns 
out that the ratios $R\equiv f/i$ are for collisional and photo-ionized plasmas comparable in the 
same density range (cf. also Mewe \cite{Mewe99}), but the ratios $G\equiv (i+f)/r$ are very different, 
e.g., $G \sim 1$ for a collisional plasma and a factor $\sim$2-4 larger for a  
photo-ionized plasma where the {\it resonance} line is relatively much weaker.

\section{Atomic data and improvements}\label{sec:improvements}

The intensities of the three  lines ({\it resonance}, {\it forbidden} and {\it intercombination}) 
are calculated mainly from atomic data presented in Paper I (Porquet \& Dubau \cite{PorquetDubau2000}). 
In this work (as well as in Paper I) for all temperatures (low and high), 
radiative recombination contributions (direct + upper--level 
radiative cascades), and collisional excitations inside the $n$=2 shell
were included in the line ratio calculations. 
For high temperatures, the collisional excitation contribution 
(direct + near-threshold resonance + cascades) from the ground level 
($n$=1 shell, 1s$^{2}$) are important as well as dielectronic recombination (direct + cascades). \\ 

Excitation collisional data are also taken from Paper I, which are based on the calculations from Zhang \& Sampson 
(\cite{Zhang87}) plus the contribution of the upper-level ($n>$2) radiative cascades calculated 
in Paper I (see Paper I for more 
details).\\ 
The ionization balance is from Mazzotta et al. (\cite{Mazzotta98}) and the data for radiative and dielectronic 
recombinations are from Paper I. Various new data for the transition probabilities 
(e.g., {\it forbidden} and  {\it intercombination} lines) have been selected (see $\S$~\ref{sec:Aki}).\\

In the following paragraphs we describe the several differences between Paper\,I and this work: 
$A_{ki}$, optical depth, 
contribution of the blended dielectronic satellite lines, and radiation field.\\

\subsection{Update of the A$_{ki}$ for the forbidden and the intercombination lines}\label{sec:Aki}

\begin{table}[!h]
\caption{Update of the transition probabilities (s$^{-1}$) with published experimental values for the {\it forbidden} line ($z$, A$_{1\to2}$) and the {\it intercombination} line ($y$, A$_{1\to4}$) compared to the theoretical 
values from Porquet \& Dubau (\cite{PorquetDubau2000}).}
\begin{tabular}{lll}
\hline
\hline
ion           & {\it forbidden} line     & {\it intercombination} line     \\
              & ($z$, A$_{1\to2}$)   & ($y$, A$_{1\to4}$)          \\
\hline
\ion{C}{v}    &  4.857\,(+1) (S94)       &  2.90\,(+7) (H85) \\               
 \ion{N}{vi}   &  2.538\,(+2) (N00)       &  1.38\,(+8) (H85) \\               
\ion{O}{vii}  &  1.046\,(+3) (C98)       &  5.800\,(+8) (E81)\\               
\ion{Ne}{ix}  &  1.09\,(+4) (T99)      &  5.400\,(+9)  *    \\                
\ion{Mg}{xi}  &  7.347\,(+4) (S95)       &  3.448\,(+10) (A81)     \\                
\ion{Si}{xiii} &  3.610\,(+5) *      &  1.497\,(+11) (A79)      \\                
\hline
\hline
\end{tabular}
\begin{description}
{\tiny
\item [(A79)]: Armour et al. (\cite{Armour79})
\item [(A81)]: Armour et al. (\cite{Armour81})
\item [(C98)]: Crespo L{\'o}pez-Urrutia et al. (\cite{Crespo98})
\item [(E81)]: Engstr{\"o}m et al. (\cite{Engstrom81})
\item [(H85)]: Hutton et al. (\cite{Hutton85})
\item [(N00)]: Neill et al. (\cite{Neil2000})
\item [(S94)]: Schmidt et al. (\cite{Schmidt94})
\item [(S95)]: Stefanelli et al. (\cite{Stefanelli95})
\item [(T99)]: Tr{\"a}bert et al. (\cite{Trabert99})
\item [*]: Theoretical values taken from Lin et al. (\cite{Lin77}), see text ($\S$\ref{sec:Aki}).
}
\end{description}
\label{table:Aki-updated}
\end{table}

We have updated the transition probabilities A$_{ki}$ reported in Paper\,I  
for the {\it intercombination} ($y$, A$_{1\to4}$) and the {\it forbidden} ($z$, A$_{1\to2}$) line by published experimental values (see Table~\ref{table:Aki-updated}, and references therein). In some cases, no published 
experimental values has been found and then we used the theoretical values from Lin et al. (\cite{Lin77}). 
Indeed, comparisons of their theoretical values with the experimental values reported in 
Table~\ref{table:Aki-updated} seem to show good agreement in other cases.\\
For \ion{C}{v}, the ratio $R$ is reduced by about 20$\%$ comparing the calculations using the values of 
A$_{ki}$  reported in Paper\,I, while for \ion{N}{vi} the reduction is 
less than 10$\%$. For \ion{O}{vii}, \ion{Ne}{ix}, \ion{Mg}{xi}, and \ion{Si}{xiii}, the differences
 between the current calculations using these new values of A$_{ki}$ and those reported in Paper\,I are negligible.\\

\subsection{Influence of the optical depth (resonant scattering)}

Schrijver et al. (\cite{Schrijver95}) and Mewe et al. (\cite{Mewe95b}) have investigated the possibility that 
 resonance photons are scattered out of the line of sight in late-type stellar coronae (see also Acton \cite{Acton78}). 
Indeed, in this process, a {\it resonance} line photon is absorbed by an ion in the ground state and then re-emitted, 
generally in a different direction. So, the total photon intensity integrated over 4$\pi$ remains unchanged 
but the photon distribution with respect to given direction is altered. This absorption and re-emission is 
indistinguishable from scattering and depends on the geometry of the region being observed. In general, 
photons would be scattered preferentially out of the line of sight for active regions (relatively dense areas) 
and into the line of sight for the surrounding quiet Sun (less dense area), see Schmelz et al. 
(\cite{Schmelz97}) and  Mewe et al. (\cite{Mewe2001}). The effect is smaller for instruments with a 
larger field of view.\\
This could have an impact on the temperature diagnostic, the so-called $G=(z+x+y)/w$ 
or $(f+i)/r$ ratio. If the optical depth of the line is not taken into account, 
the calculated intensity ratio $G$ can be overestimated and 
 the inferred temperature from the $G$ ratio is underestimated.

As detailed in Mewe et al. (\cite{Mewe2001}), branching ratios can be used to check the assumption of the 
optical thin model because effects of resonance scattering would affect the measured branching ratio. From 
the fact that the intensities of e.g., the strong resonance lines \ion{Fe}{xviii}$\lambda$93.92 and 
\ion{Fe}{xix}$\lambda$108.307 are in good agreement with the intensities of other lines sharing the same 
upper level, one can derive a constraint on the optical depth taking into account the systematic 
uncertainties of the theoretical transition probabilities A (typical 25$\%$ for each A, hence 35$\%$ for 
the branching ratio) which dominate over the statistical errors (typically 10$\%$). If we rule out a 
reduction in the resonance line intensity larger than about 30$\%$, then on the basis of a simple 
``escape-factor'' model with
\begin{equation}\label{eq:Pr}
P(\tau) \simeq \frac{1}{[1+0.43 \tau]},
\end{equation}
the escape factor for a homogeneous mixture of emitters and absorbers in a slab geometry 
(e.g., Kaastra $\&$ Mewe \cite{KaastraMewe95}), one can put a constraint on the optical depth.
The optical depth $\tau$ for a Doppler-broadened resonance line can be written as 
(Mewe et al. \cite{Mewe95b})\footnote{
the forefactor in the first part of Eq. (23) was in Mewe et al. (\cite{Mewe95b},\cite{Mewe2001}) 
erroneously taken a factor of 10$^3$ too large, but the 2nd part of Eq. (23) is still correct.}:   
\begin{eqnarray}\label{formula-Mewe95b}
\tau&=&1.16\,10^{-17}~\left(\frac{n_{i}}{n_{el}}\right)A_{Z}\left(\frac{N_{\rm {H}}}{n_{\mathrm e}}\right)\lambda f \sqrt{\frac{M}{T_{\mathrm e}}}  n_{\mathrm e} l\nonumber\\
    &\equiv&10^{-19}~C_{d}~\left(\frac{A_{Z}}{A_{Z_{\odot}}}\right)~\left(\frac{n_{\mathrm e}~l}{\sqrt{T_{6}}}\right), 
\end{eqnarray}
where ($n_{i}/n_{el}$) is the ion fraction (e.g. from Arnaud \& Rothenflug \cite{Arnaud85}, from 
Arnaud \& Raymond \cite{Arnaud92} for iron, or from Mazzotta et al. \cite{Mazzotta98}), A$_{Z}$ is the elemental abundance relative to hydrogen, 
$A_{Z_{\odot}}$ the corresponding value for the solar photosphere as given by Anders \& Grevesse 
(\cite{Anders89}), $N_{\rm H}/n_{\rm e}\simeq$0.85 the ratio of hydrogen to electron density, $\lambda$ is
 the wavelength in \AA, $f$ the absorption oscillator strength,  $M$ is the atomic weight, $T_{\mathrm e}$ is the 
 electron temperature (in K or $T_{6}$ in MK), $l$ a characteristic plasma dimension (in cm) and 
\begin{eqnarray}
C_{d}\equiv 98.5 \left(\frac{n_{i}}{n_{el}}\right) A_{Z_{\odot}} \lambda f  \sqrt{M} .
\end{eqnarray}

According to  Equation (\ref{formula-Mewe95b}), 
Ness et al. (\cite{Ness2001a}) estimated the optical depth,
adopting a value of unity for the fractional ionization and using solar abundances. One further assumes
 $T_{\mathrm e}$ at the peak line formation, but note that $\tau$ is rather insensitive to the precise value of 
 $T_{\mathrm e}$. 
 One can determine -- for each {\it resonance} line -- that value of $n_{\mathrm e} \ell$ which yields an 
 optical depth of 
 unity. According to the values of n$_{e}$ inferred from the ratio $R=z/(x+y)$ or $R=f/i$ (from \ion{C}{v} 
 to \ion{Si}{xiii} 
the $intercombination$ and the  $forbidden$ lines are not sensitive to resonant scattering 
below a column density of ${\cal N}_{\rm H}\sim 10^{25-26}$~cm$^{-2}$
  and ${\cal N}_{\rm H}\sim 10^{30-31}$~cm$^{-2}$, respectively, while the {resonance} line becomes sensitive to the 
resonant scattering above ${\cal N}_{\rm H}\sim 10^{21-23}$~cm$^{-2}$), one can determine the corresponding values of 
  $\ell$. One  can compute the respective emission measures of $n_{\mathrm e}^2\ell^{\,3}$ respectively, and can 
  compare these emission measures with those derived from the measured line fluxes $f_\lambda$ according to
\begin{equation}
EM=\frac{4\pi d^{\, 2} f_{_{\lambda}}}{P_{_{\lambda}}(T_{\rm e})}
\end{equation}
with the line emissivity $P_\lambda(T_{\rm e})$ and the distance $d$ of the star. If  the former is larger, this 
inconsistency shows that the assumption of a non-negligible optical depth is invalid and we conclude that 
optical depth effects are irrelevant for the analysis of He-like triplets. On the contrary the effect of 
resonant scattering should be taken into account when comparing the theoretical values with the observational 
ones.\\

Since $\tau_{_{r}}>>\tau_{_{i}}$ ($\tau_{_{r}}$ and $\tau_{_{i}}$ corresponding respectively to the optical depth of 
the {\it resonance} and the {\it intercombination} lines), we can write $G_{\tau}\equiv \frac{G}{P_{_{r}}}$,
where $G_{\tau}$ is the value of the ratio taken into account the optical depth of the {\it resonance} line, 
$G$ is the value without  resonant scattering (such as in Paper\,I and $\S$\ref{sec:results}), and 
$P_{_{r}}$ is the escape probability for the {\it resonance} line (Eq.~\ref{eq:Pr}). 
One should note that $G_{\tau}$ is not strictly exact when the contribution of the blended dielectronic satellite lines are 
introduced in the calculations (see $\S$\ref{sec:satlines}).\\

\subsection{Blended dielectronic satellite lines }\label{sec:satlines}

The intensity of a dielectronic satellite line arising from a doubly excited state with principal quantum number $n$ 
in a Lithium-like ion produced by dielectronic recombination of a He-like ion  is given by: 
\begin{equation}
I_{s}=N_{\rm He}~ n_{\mathrm e}~ C_{s},
\end{equation}
where $N_{\rm He}$ is the population density of the considered He-like ion in the ground state 1s$^{2}$ with statistical
weight $g_1$ (for He-like ions $g_1=1$).\\
 The rate coefficient (in cm$^{3}$\,s$^{-1}$) for dielectronic recombination is given by (Bely-Dubau et al. 
 \cite{Bely-Dubau79}):
\begin{equation}
C_{s}=2.0706\ 10^{-16}~\frac{e^{-E_{s}/kT_{\mathrm e}}}{g_{1} T_{\mathrm e}^{3/2}}~F_{2}(s),
\end{equation}
where $E_{s}$ is the energy of the satellite level $s$ with statistical weight $g_s$ above the  
ground state of the He-like ion, $T_{\mathrm e}$ is the electron temperature in K, and 
$F_{2}(s)$ is the so-called line strength factor (often of the order of about 10$^{13}$~s$^{-1}$ for the stronger lines) given by
\begin{equation}
F_{2}(s) = {{g_s A_a A_r} \over {(A_a + \sum A_r)}},
\end{equation}
where $A_a$ and
$A_r$ are transition probabilities (s$^{-1}$) by autoionization and radiation,
and the summation is over all possible radiative transitions from the satellite level $s$. \\
For a group of satellites with the same principal quantum
number $n$, $E_s$  can be approximated by
\begin{equation}\label{eq:Es}
E_s [\rm eV] = 1.239842\ 10^4 \ {a_{DR}\over {\lambda}},
\end{equation}
where $\lambda$ is the wavelength (\AA) of the satellite line and $a_{DR} \simeq$  
0.7, 0.86, 0.92, and 0.96 for $n$ = 2, 3, 4, and $>$ ~4, respectively
(Mewe \& Gronenschild \cite{Mewe81}).
For $\lambda$ in \AA\ and $T$ in K we can write:
\begin{equation}
{E_s\over kT} = {a_{DR} hc\over {\lambda kT}} = 1.439\ 10^8 \ {a_{DR}\over { \lambda T}}.
\end{equation}

The influence of the blending of dielectronic satellite lines for the {\it resonance}, the {\it intercombination} and the 
{\it forbidden} lines has been taken into account where their contribution is not negligible in the 
calculation of $R$ and $G$, affecting the inferred electron temperature and density. This is the case 
for the high-Z ions, i.e. \ion{Ne}{ix}, \ion{Mg}{xi}, and \ion{Si}{xiii} (Z=10, 12, and 14, respectively).
 Since the contribution of the blended dielectronic satellite lines depends on the spectral resolution
 considered, we have estimated the ratios $R$ and $G$ for four specific spectral resolutions (FWHM): 
RGS-1 at the first order 
(i.e. $\Delta \lambda$=0.073, 0.075 and 0.078\AA~ for \ion{Ne}{ix}, \ion{Mg}{xi} and \ion{Si}{xiii} respectively),
LETGS (i.e. $\Delta \lambda$=0.05\AA), HETGS-MEG (i.e. $\Delta \lambda$=0.023\AA), and HETGS-HEG 
(i.e. $\Delta \lambda$=0.012\AA).

For the $n$=2, 3, 4 blended dielectronic satellite lines we use the atomic data reported in the appendix. 
For the higher-n blended dielectronic satellite lines we use the results from Karim and co-workers.
For Z=10 (\ion{Ne}{ix}) we use the data from Karim (\cite{Karim93}) who gives the intensity factor 
$F^{*}_{2} \equiv F_2/g_1$ for 
the strongest ($F^{*}_{2} > 10^{12}$\,s$^{-1}$) dielectronic satellite lines with $n$=5-8. For Z=14 (\ion{Si}{xiii}), 
we take the calculations from Karim \& Bhalla (\cite{Karim92}) who report the intensity factor 
$F^{*}_{2}$ for the strongest ($F^{*}_{2} > 10^{12}$\,s$^{-1}$) dielectronic satellite lines 
with $n$=5-8. For Z=12 (\ion{Mg}{xi}) we have interpolated between the calculations from Karim (\cite{Karim93}) for 
Z=10, and from 
Karim \& Bhalla  (\cite{Karim92}) for Z=14.\\
Including the contribution of the blended dielectronic satellite lines, we write for the ratios $R$ and $G$:
\begin{equation}
R=\frac{z+satz}{(x+y)+satxy}
\end{equation}
\begin{equation}
G=\frac{(z+satz)+((x+y)+satxy)}{(w+satw)},
\end{equation}
where $satz$, $satxy$ and $satz$ are respectively the contribution of blended dielectronic satellite lines to the 
{\it forbidden} line, to the {\it intercombination} lines, and to the {\it resonance} line, respectively. 
One can note that at very high density the $^{3}$P levels are depleted to the $^{1}$P level, and in that 
case {\it x+y} decreases and $R$ tends to $satz$/$satxy$.\\

At the temperature at which the ion fraction is maximum for the He-like ion 
(see e.g. Arnaud \& Rothenflug \cite{Arnaud85}, Mazzotta et al. \cite{Mazzotta98}), the differences between the 
calculations for $R$ (for $G$) with or without taking into account the blended dielectronic satellite lines are only
of about 1$\%$ (9$\%$), $2\%$ (5$\%$), and 5$\%$ (3$\%$) for \ion{Ne}{ix}, \ion{Mg}{xi}, and 
\ion{Si}{xiii} at the low-density limit and for $T_{\mathrm rad}$=0\,K, respectively. On the other hand, for much 
lower electron temperatures, the effect is bigger since the intensity of 
the dielectronic satellite lines is proportional to T$_{e}^{-3/2}$. As well, for high values of density 
($n_{\mathrm e}$) at which the intensity of the {\it forbidden} line is very weak (i.e. tends to zero), the 
contribution of the blended dielectronic satellite lines to the {\it forbidden} line leads to a ratio $R$ which decreases much 
slower with $n_{\mathrm e}$ than in the case where the contribution of the blended dielectronic satellite lines is not taken 
into account.

\subsection{Influence of a radiation field}\label{sec:photo-excitation}

Recently, Kahn et al. (\cite{Kahn2001}) have found with the {\sl RGS} on {\sl XMM-Newton} 
that for $\zeta$ Puppis, the $forbidden$ to 
$intercombination$ line ratios within the Helium-like triplets are abnormally low for \ion{N}{vi}, 
\ion{O}{vii}, and \ion{Ne}{ix}. While this is sometimes indicative of a high electron density, 
they have shown that 
in the case of $\zeta$ Puppis, it is instead caused by the intense radiation field of this star. This constrains 
the location of the X-ray emitting shocks relative to the star, since the emitting regions should be close enough 
to the star in order that the UV radiation is not too much diluted. 
\\  
A strong radiation field can mimic a high density if the upper ($^{3}$S) level of the {\it forbidden} line 
is significantly depopulated via photo-excitation to the upper ($^{3}$P) levels 
of the {\it intercombination} lines, 
analogously to the effect of electronic collisional excitation (fig.~\ref{fig:gotrian}). 
The result is an increase of the {\it intercombination} 
lines and a decrease of the {\it forbidden} line.
\\
Eq. (21) gives the expression for photo-excitation from level $m$ to level $p_k$ in a radiation field
with effective blackbody temperature $T_{\mathrm rad}$ from a hot star underlying the X-ray line emitting plasma. As pointed
out by Mewe  \& Schrijver (\cite{Mewe78a}) the radiation is diluted by a factor $W$ given by
\begin{equation}
W=\frac{1}{2}~\left[1-\left(1-\left(\frac{r_{*}}{r}\right)\right)^{1/2}\right],
\end{equation}
where $r$ is the distance from the center of the stellar source of radius $r_{*}$. 
Close to the stellar surface the dilution factor $W={1\over 2}$.
For stars such as Capella or Procyon, we can take $W={1\over 2}$, because the stellar 
surface which is the origin of the radiation irradiates coronal structures that are close to
the stellar surface (Ness et al. \cite{Ness2001a}). In a star such as Algol the 
radiation originates from another star, and $W$ is much lower (i.e. $W \simeq 0.01$, cf. Ness et al. \cite{Ness2001b}),
but due to the strong radiation field the radiation effect can still be important.\\ 

In their Table~8,  Mewe \& Schrijver (\cite{Mewe78a}) give for information the radiation temperature for a 
solar photospheric field for Z=6, 7, and 8. In Table~\ref{table:Trad}, we report the wavelengths at 
which the radiation temperature should be estimated  for Z=6, 7, 8, 10, 12, 14. 
These wavelengths correspond to the transitions between the $^3$S and $^3$P levels ($\lambda_{f\to i}$) and the 
$^1$S and $^1$P levels ($\lambda_{6\to r}$).\\

\begin{table}[!h]
\caption{Wavelengths at which the radiation temperature (T$_{rad}$) should be determined.}
\label{table:Trad}
\begin{tabular}{ccccccc}
\hline
                       &  \ion{C}{v}     &    \ion{N}{vi}     &   \ion{O}{vii}    &  \ion{Ne}{ix}   & \ion{Mg}{xi}   &  \ion{Si}{xiii} \\
$\lambda_{f\to i}$ (\AA)& 2280            & 1906               &  1637             & 1270            &  1033          & 864 \\
$\lambda_{6\to r}$ (\AA)& 3542            & 2904               &  2454             & 1860            &  1475          & 1200\\
\hline
\end{tabular}
\end{table}

 The photo-excitation from the  $^{3}$S level and  $^{3}$P levels is very important for low-Z ions \ion{C}{v}, 
 \ion{N}{vi}, \ion{O}{vii}. For higher-Z ions, this process is only important for very high radiation temperature 
 (~$\sim$few 10\,000\,K).\\

One can note that the photo-excitation between the levels $^{1}$S$_{0}$ and $^{1}$P$_{1}$ is negligible compared 
to the photo-excitation between the  $^{3}$S$_{1}$ and $^{3}$P$_{0,1,2}$ levels. For example, for a very high  
value of T$_{\mathrm rad}$=30\,000\,K the difference between the calculations taken or not taken into account the 
photo-excitation between $^{1}$S$_{0}$ and $^{1}$P$_{1}$ is smaller than 20$\%$ for \ion{C}{v}, where this effect 
is expected to be maximum.

\section{Results from extended calculations}\label{sec:results}

Using the above-mentioned atomic data, we have calculated the line intensity ratios $R$ and $G$ 
for \ion{C}{v}, \ion{N}{vi}, \ion{O}{vii}, \ion{Ne}{ix}, \ion{Mg}{xi}, 
and \ion{Si}{xiii}\footnote{Tables 4 to 69 are only available in electronic form
 at the CDS via anonymous ftp to cdsarc.u-strasbg.fr (130.79.128.5)
 or via http://cdsweb.u-strasbg.fr/cgi-bin/qcat?J/A+A/.}
The wavelengths of these three (four) lines for each  He-like ion treated in this paper 
are reported in Table~\ref{table:lambda}.\\
All the relevant processes detailed in sections~\ref{sec:schematic-model} and \ref{sec:improvements} 
between the seven levels are taken into account (full resolution): 
radiative de-excitation, collisional electronic excitation and de-excitation, 
radiative and dielectronic recombination, photo-excitation and induced emission 
(between $^{3}$S$_{1}$ and $^{3}$P levels). \\
We considered a broad range of densities ($n_{\mathrm e}$) and radiation temperatures ($T_{\mathrm rad}$) calculated 
for photo-excitation between the $^3$S level and the $^3$P levels and a number of electron temperatures ($T_{\mathrm e}$).
As well, we considered different values of the dilution factor of the radiation field ($W$) 
which could be used either for hot late-type stars or O, B stars.\\

We display the $G(T_{\mathrm e})$ line intensity ratios, from Tables~4 to 9, 
for the six ions, for five values 
of electron temperature ($T_{\mathrm e}$) including the temperature of maximum line formation for the He-like 
lines  (cf. Mewe et al. \cite{Mewe85}); and for two or more values of the radiation temperature ($T_{\mathrm rad}$), and 
several values of $n_{\mathrm e}$. As one can note the ratio $G$ is as expected sensitive to $T_{\mathrm e}$, while 
it is almost insensitive to the exact values of $n_{\mathrm e}$ and $T_{\mathrm rad}$. 
The {\it resonance} line becomes sensitive at high density due to the depopulation of the 
1s2s$^{1}$S$_{0}$ level to the 1s2p$^{1}$P$_{1}$ level (see Gabriel \& Jordan \cite{GabrielJordan72}). 
Since the sum $z+(x+y)$ is a  constant or almost constant, 
the value of $G$ is independent of the exact value of the dilution factor ($W$). Here the calculations were done for dilution factor $W$=1/2.

Finally, we display the $R(n_{\mathrm e})$ line intensity ratios for the six ions, 
in Tables~10 to 69, 
for the same values of electron temperature ($T_{\mathrm e}$) and much more values of radiation temperature 
($T_{\mathrm rad}$), and for three values of the dilution factor of the radiation field $W$=0.5, 0.1, 0.01.\\

Since as said previously in $\S$\ref{sec:satlines}, the contribution of the blended dielectronic satellite lines depends 
on the spectral resolution, we give the values of $R$ and $G$ for \ion{Ne}{ix}, \ion{Mg}{xi}, and \ion{Si}{xiii}, for four specific values of spectral resolutions (FWHM):
 RGS-1 at the first order 
(i.e. $\Delta \lambda$=0.073, 0.075 and 0.078\AA~ for \ion{Ne}{ix}, \ion{Mg}{xi} and \ion{Si}{xiii} respectively),
LETGS (i.e. $\Delta \lambda$=0.05\AA), HETGS-MEG (i.e. $\Delta \lambda$=0.023\AA), and HETGS-HEG 
(i.e. $\Delta \lambda$=0.012\AA). In the cases where the differences ($<$1$\%$) are negligible 
between two or more spectral resolutions, we display the results together (e.g. RGS and LETGS). 

\section{Conclusions}

For the first time, thanks to the new generation of X-ray satellites, {\sl Chandra} and {\sl XMM-Newton} the diagnostics based on the line ratios of He-like ions could be used for powerful extra-solar plasma diagnostics (Warm Absorber in AGNs, stellar coronae, ...). These diagnostics are one of the keys for a better understanding of the solar-stellar connection: heating of the coronae, magnetic activity, etc. 
In this work, we have calculated the line ratio $R$ and $G$ which allow, respectively electron density and temperature diagnostics. These calculations are based partly on the previous work of Porquet \& Dubau (\cite{PorquetDubau2000}) and on different improvements of atomic data (see $\S$\ref{sec:improvements}).\\
We have tabulated the results adapted for the different spectral resolutions of the spectrometers on board of {\sl Chandra} and {\sl XMM-Newton} for parameter ranges which correspond as much as possible to  most cases of stellar coronae (hot-late type star, O B stars...).

\begin{acknowledgements}
The Space Research Organization Netherlands (SRON) is supported financially by NWO.
\end{acknowledgements}

\newpage
\appendix
\section{Atomic data for the blended dielectronic satellite lines with $n$=2, 3, 4.}

We report in this appendix the atomic data related to the  dielectronic satellite lines, 
 calculated in this work for Z=10, 12, and 14, blended with one of their parent lines:
{\it forbidden}, {\it intercombination}, and {\it resonance}.\\
The satellite line wavelengths and intensities have been obtained using
a set of computer codes developed partly in University-College-London 
(SUPERSTRUCTURE: Eissner, Jones, Nussbaumer \cite{Eissner74}) and in Meudon Observatory 
(AUTOLSJ: TFR group, Dubau J., Loulergue M. \cite{TFR81}). Multiconfigurational-wavefunctions are 
calculated in a "scaled" Thomas-Fermi-Dirac-Amaldi potentials, depending 
on linear scaling parameters ($\lambda_{s}$, $\lambda_{p}$,$\lambda_{d}$ ...) different 
for $l$-orbitals, which are obtained through a self-consistent variational
procedure on the energy sum of the first lowest (SL) terms. In 
SUPERSTRUCTURE, the level energies and the radiative probabilities are 
calculated in the relativistic Breit-Paul hamiltonian approach, which gives 
fine-structure bound and autoionizing levels. In the AUTOLSJ code, the 
autoionization probabilities are derived in the Distorted-wave 
approximation, using the same wavefunctions as in SUPERSTRUCTURE.
For the present calculations, the following configuration were used :
$1s^2nl$, $1s2snl$, and $1s2pnl$ for $n = 2,\cdots,5$ and $0\le l \le n-1$.\\

The wavelengths of the dielectronic satellite lines calculated here should be 
compared to the ``reference'' wavelengths used in the Jacques Dubau's calculations 
respectively for \ion{Ne}{ix}, \ion{Mg}{xi}, and \ion{Si}{xiii}, 
$\lambda_{w}$=13.4658, 9.1740, 6.6482\,\AA, $\lambda_{y}$=13.5774, 9.2395, 6.6903\,\AA, 
$\lambda_{x}$=13.5774, 9.2358, 6.6865\,\AA, $\lambda_{z}$=13.7216, 9.3219, 6.7420\,\AA. 
One can notice that these wavelengths are not identical to the wavelengths of 
Vainshtein \& Safronova (\cite{Vainshtein78}) used in the calculation 
of the line ratios $R$ and $G$, tabulated in Table~\ref{table:lambda}. 
Then in order to determine which dielectronic satellite lines are blended 
with one of the parent lines ({\it forbidden}, {\it intercombination}, 
and {\it resonance}), one should take into account the shift of the 
satellite line compared to the wavelengths chosen for the parent lines 
in the calculation of $R$ and $G$.\\
The values of E$_{s}$, which is the energy of the satellite level $s$, 
used in this calculation are well reproduced using formula (\ref{eq:Es}). \\

In tables\footnote{Tables A.1 to A.6 are only available in electronic form.}~A.1, 
~A.2, and ~A.3, 
the dielectronic satellite lines $n$=2, for Z=10, 12, and 14, respectively 
are reported. In tables~A.4, ~A.5, and ~A.6, the dielectronic satellite lines $n$=3, 
and 4, for Z=10, 12, and 14, respectively are reported.

\end{document}